\documentclass[12pt]{iopart}

\usepackage{graphicx}
\begin{document}

\title{Spin configuration in a frustrated ferromagnetic/antiferromagnetic thin film system}
\author{T K Yamada,$^{\rm 1}$ E Mart\'{\i}nez,$^{\rm 2}$ A Vega,$^{\rm 3}$ R Robles,$^{\rm 4}$ D Stoeffler,$^{\rm 5}$ A L  V\'azquez de Parga,$^{\rm 6}$ T Mizoguchi,$^{\rm 1}$ and H van Kempen$^{\rm 7}$}
\address{$^{\rm 1}$Faculty of Science, Gakushuin University,
171-8588 Mejiro, Tokyo, Japan} 
\address{$^{\rm 2}$Fachbereich Physik, Universit\"at Osnabr\"uck, D-49069 Osnabr\"uck, Germany}
\address{$^{\rm 3}$Dep. de
F\'{\i}sica Te\'orica, At\'omica y \'{O}ptica, Universidad de
Valladolid, 47011 Valladolid, Spain} 
\address{$^{\rm 4}$Department
of Physics, Uppsala University, SE-75121, Sweden}
\address{$^{\rm 5}$Institut de Physique et Chimie des Mat\'eriaux
de Strasbourg (UMR C7504 CNRS-ULP), Strasbourg, France}
\address{$^{\rm 6}$Dep. F\'{\i}sica de la Materia Condensada,
Universidad Aut\'onoma de Madrid, Cantoblanco 28049, Madrid,
Spain} 
\address{$^{\rm 7}$Institute for Molecules and Materials,
Radboud University, Toernooiveld 1, 6525 ED Nijmegen, The
Netherlands}

\ead{edmartin@uos.de}
\begin{abstract}
We have studied the magnetic configuration in ultrathin antiferromagnetic Mn
films grown around monoatomic steps on an Fe(001) surface by
spin-polarized scanning tunneling microscopy/spectroscopy and
\emph{ab-initio}-parametrized self-consistent real-space tight
binding calculations in which the spin quantization axis is
independent for each site thus allowing noncollinear magnetism. Mn
grown on Fe(001) presents a layered antiferromagnetic structure.
In the regions where the Mn films overgrows Fe steps the
magnetization of the surface layer is reversed across the steps.
Around these defects a frustration of the antiferromagnetic order
occurs. Due to the weakened magnetic coupling at the central Mn
layers, the amount of frustration is smaller than in Cr and the
width of the wall induced by the step does not change with the
thickness, at least for coverages up to seven monolayers.
\end{abstract}

\maketitle
\section{Introduction}

The interface of ferromagnetic and antiferromagnetic material is
important from a scientific point of view because the competing
ferromagnetic and antiferromagnetic interactions may lead to
complex configurations particularly when there is frustration in the
system. Especially when an antiferromagnetic layer is deposited on
a ferromagnetic substrate with an atomic step ("hidden atomic
step") the magnetic frustration around this extended defect can
give rise to interesting magnetic structures \cite{Berger, Vega,
Stoeffler, Berger1}. Due to the localized nature of the
frustrations, it has not been possible to resolve the spin
configurations until the introduction of the Spin-Polarized
Scanning Tunneling Microscopy/Spectroscopy (SP-STM/STS)
\cite{Bode}.

This problem is also important from the technological point of
view. Exchange bias is one of the phenomena 
associated with the exchange anisotropy created 
at the interface between an antiferromagnetic 
material and a ferromagnetic material. Materials 
exhibiting exchange bias have been used in 
several practical applications since their 
discovery \cite{Prinz,exchangebias,Wolf}. In the world of magnetic devices, the
goal is to get smaller. The smaller space one bit of information
can occupy, the more data you can get into a device. Between two
magnetic domains with opposite magnetization directions always
exists a domain wall. Therefore a deep understanding of the
parameters that control the domain wall width are crucial in order
to achieve higher density for data storage.

Mn is exactly in the middle of the 3$d$ transition metal series,
just between Fe, which is a natural ferromagnet in the bulk, and
Cr, which is an antiferromagnet. Therefore, Mn stands as one of
the more complex 3$d$ transition metals from the point of view of
the magnetic coupling, and it is a clear candidate to exhibit a
great variety of magnetic structures. Mn systems have been
experimentally investigated by spin-polarized electron energy loss
spectroscopy \cite{Walker}, scanning electron microscopy with
polarization analysis \cite{Tulchinsky} and SP-STM/STS
\cite{toyoPRL,schlickum}. A layered antiferromagnetic (LAF)
arrangement was found in the Mn film. Recently, part of the
authors \cite{physletA} have studied the same system using the
\emph{ab-initio} tight-binding linear muffin-tin orbitals
(TB-LMTO) method \cite{andersen}, assuming the experimental
interlayer distances and a $p(1\times 1)$ magnetic arrangement at
the surface as experimentally observed \cite{toyoPRL}. Different
magnetic solutions with energy differences of few meV were
obtained, with the LAF configuration the less energetic state, in
good agreement with the experiments. The LAF as well as the
closest metastable solutions had some common features: i) parallel
coupling at the interface between Mn and Fe, ii) antiparallel
coupling between the Mn-surface and subsurface layers and iii)
antiparallel coupling between the two Mn layers closest to the
interface. This set of solutions only differ in the couplings at
the central Mn layers, that were parallel or antiparallel
depending on the Mn thickness, which make these systems clear
candidates to exhibit noncollinear magnetic arrangements under
structural defects like monoatomic steps. Theoretically, Hafner
and Spi\v{s}\'ak \cite{hafner} have found complex magnetic
coupling in Mn films on Fe(001) allowing atomic relaxations,
obtaining an in-plane antiferromagnetic structure.

In this paper, the magnetic structure of thin Mn films grown on
Fe(001) is studied. In particular, we focus our attention at the
magnetic structure of the Mn films around steps on the Fe(001)
substrate. In Section 2, by means of SP-STM/STS, we measure in real space and
with high spatial resolution the magnetic structure of the films
around these defects. The experimental results are interpreted, in Section 3,
with the help of \emph{ab-initio}-parametrized self-consistent
real-space tight binding (TB) calculations in which the spin
quantization axis is independent for each site thus allowing
noncollinear magnetism. Throughout the paper, comparisons are made
with the Cr/Fe(001) system to get a deeper understanding of which
material parameters are crucial to determine the resulting
magnetic structure. Finally, in Sec. 4 we summarize our conclusions.

\section{Experimental results}
All measurements were performed in an ultra-high vacuum (UHV)
chamber (${\rm \sim5\times10^{-11}}$ mbar) at room temperature
(RT). An STM was attached to the chamber, which is equipped with
molecular beam epitaxy, Auger spectroscopy, field emission
spectroscopy, sample heating and sputtering facilities
\cite{toyoPRL}. W tips were cleaned by Ar sputtering and annealing
and then covered with 10 nm Fe. SP-STS measurements were performed
in two different ways: In the first one, $I(V)$ curves were
obtained at every pixel of a constant current topographic image
and then numerically differentiated. In the second one, $dI/dV$
maps were obtained with a lock-in amplifier modulating the sample
bias voltage by 40mV at 2 kHz.

\begin{figure}
\begin{center}
\includegraphics*[width=84mm]{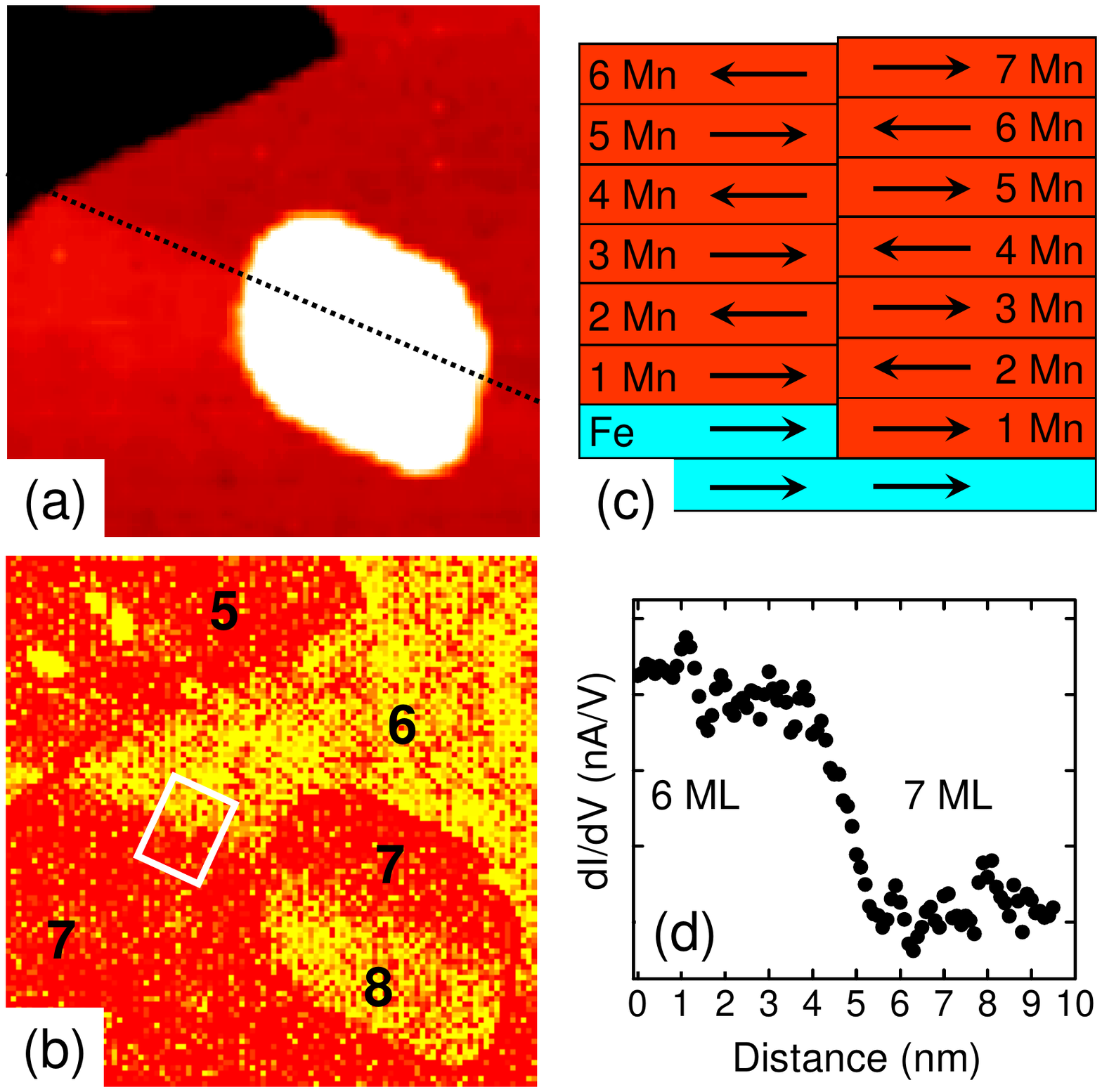}
\caption{(color online) (a) STM image after deposition of 5 Mn
ML's (70$\times$70nm$^2$). The dotted line marks the position of
the hidden step. (b) $dI/dV$ map taken at +0.2 V. The numbers
indicate the Mn local thickness. (c) Model of the Mn structure
around the hidden step. (d) Average experimental
profile across the hidden step measured inside the white box shown
in (b).}
\label{figure1}
\end{center}
\end{figure}

The growth conditions are very important because 
Mn and Fe tend to intermix and the magnetic 
properties of the Mn are very sensitive to the 
atomic structure \cite{Wu}. In this work, Mn was grown 
on the Fe(001) whisker at 100 C at a rate of 
0.6nm/min, and the surface topography and 
electronic structure was characterize by means of 
scanning tunneling microscopy/spectroscopy 
(STM/STS) at RT in UHV using a clean W tip. 
Atomically and chemically resolved STM images 
show that the Mn film grows with the same 
in-plane lattice constant as Fe(001) and that the 
Fe atoms intermix with the first, the second, and 
the third Mn layer. The concentration of 
intermixed Fe decreased with film thickness. 
Furthermore, STM shows that the growth mode 
changes from layer-by-layer to layer-plus-islands 
for coverages higher than 4 ML of Mn. If the 
temperature of the substrate is increased by few 
degrees, this transition from layer-by-layer to 
layer-plus-islands takes place for thicker Mn 
layers and  the intermixed Fe is present in more 
Mn layers. Based on apparent step height 
measurements, done choosing the bias voltage 
carefully to avoid the influence of the 
electronic structure on the results, we conclude 
that the first two Mn overlayers show a 
significant higher step heights than the Fe 
single step and that the Mn film relaxes by about 
0.02 nm at the third layer. From the fourth layer 
the interlayer spacings are geometrically the 
same (about 0.165 nm) and the structure is a 
body-centered tetragonal structure \cite{toyoSS}. This bct 
Mn(001) has a layered antiferromagnetic arrangement \cite{Tulchinsky,toyoPRL}.

When a Mn film is grown across an Fe step (hidden), the Mn film
tends to produce a flat surface. We always find steps 0.02 nm
high, that corresponds to the difference between the interlayer
distances for Fe and Mn. Across one of these steps, the Mn
thickness changes by one layer (see of Fig. \ref{figure1}(c)). Due
to the LAF structure, the magnetization of the surface layer is
reversed across these steps. $dI/dV$ maps obtained with clean W
tips around these defects show the same electronic structure on
both sides. However, when using Fe covered W tips, contrast of
magnetic origin is obtained across the hidden step as can be seen
in Fig. \ref{figure1}(b). The average profile measured in the
white box shown in panel (b) can be seen in panel (d) and gives a
domain wall width of around 1.16 nm (2-4 lattice parameters).

\begin{figure}
\begin{center}
\includegraphics*[width=120mm]{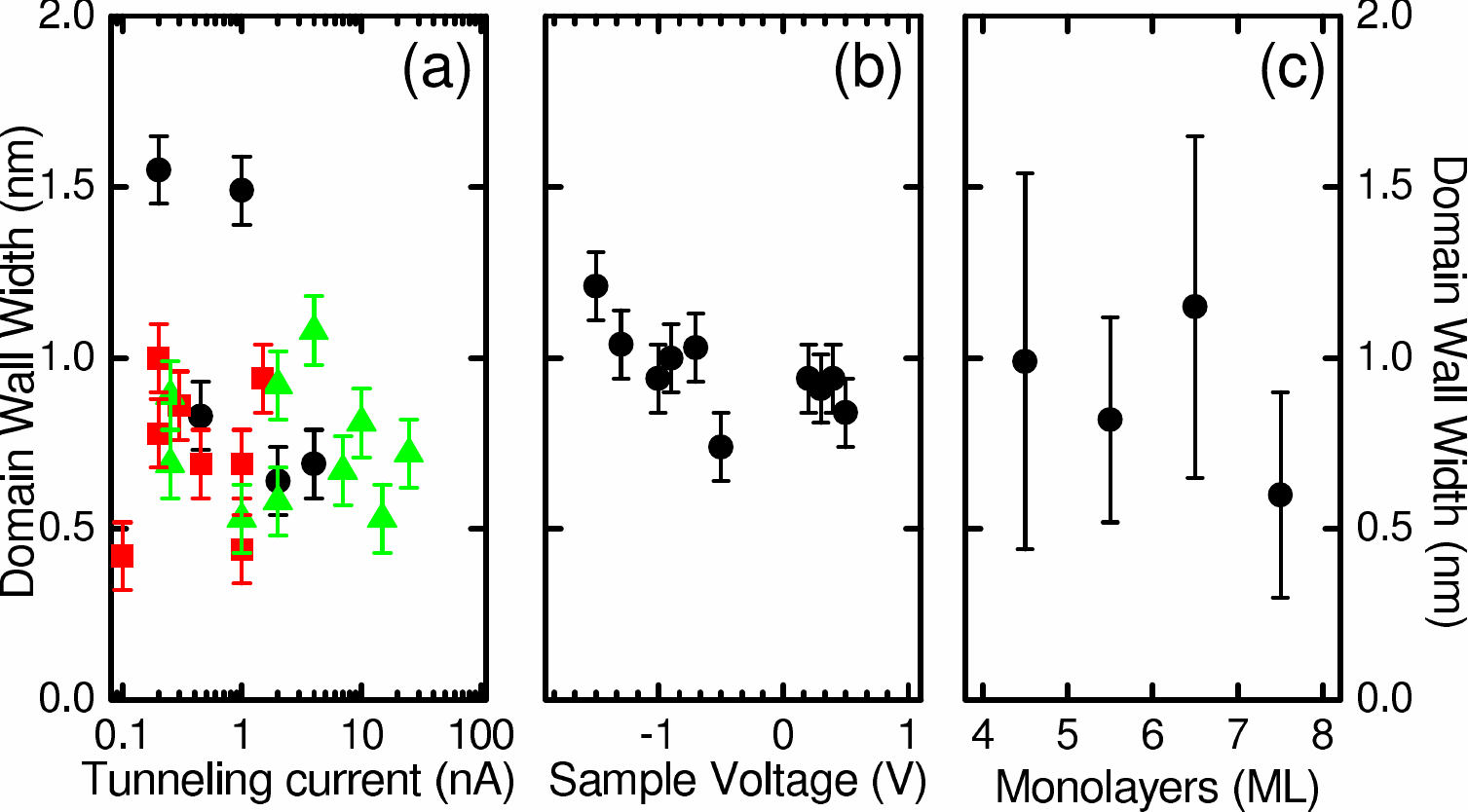}
\caption{(color online)(a) Shows the domain wall width versus the
tunneling current (tip-sample separation). Black dots, green
triangles and red squares were obtained at a voltage set point of
V$_s$ = +1.5 V, V$_s$ = +0.8 V, and V$_s$ = -1.0 V, respectively.
(b) Shows the domain wall width versus bias voltage.  The
experiments were done on hidden steps between 4-5 ML and 6-7 ML.
(c) Domain wall width versus Mn thickness.} \label{figure2}
\end{center}
\end{figure}

To study the possible influence of the set-point values on the
observed domain width, we changed systematically the set point
voltage and set point current. Firstly, the set point currents
were varied with fixed set point voltage, so with varying
tip-sample distance (Fig. \ref{figure2}(a)). In addition, the set
point voltage and current were varied such that the tunneling
resistance stayed constant, so with approximately constant
tip-sample separation (Fig. \ref{figure2}(b)). From the results
given in Fig. \ref{figure2}(a) and \ref{figure2}(b), we can
conclude that our values for the domain wall width do not depend
on the particular set-point used.

In Fig. \ref{figure2}(c) we show the result of 40 measurements
taken with 6 different magnetic tips for hidden steps covered by
different Mn thicknesses. There is no clear dependence between the
width of the domain wall and the thickness of the Mn film in the
range explored. The error bars in these measurements simply
reflect the fact that the tips are different and the resolution of
the magnetic images is affected.

\section{Theoretical results}
For the theoretical investigation of our samples, it is important
to point out that when defects are present like in this system,
with the consequent lack of symmetry and hundreds of inequivalent
sites, real space methods with reasonable computer requirements
are demanded. We have employed an \emph{ab-initio}-parametrized
self-consistent real-space tight-binding method in which the spin
quantization axis is independent for each site thus allowing
noncollinear magnetism. This method has been recently developed
and used satisfactory for the study of supported Cr films on Fe
substrates with monoatomic steps \cite{nocolcr}. The Hamiltonian
in our method can be split into a band term $H_{band}$ and an
exchange term $H_{exch}$, which in the orthogonal $|i\alpha
\rangle$ basis of atomic site $i$ and orbital $\alpha$ and with
the usual notation are:

\begin{equation}
\nonumber H=H_{band}+H_{exch}
\end{equation}

%\begin{equation}
\begin{eqnarray}
\nonumber H_{band}=\sum_{\scriptstyle i,j \atop \scriptstyle
\alpha,\beta}[(\epsilon ^{0}_{i\alpha } + U_{i\alpha,j\beta}
\langle\hat{n}_{jb}\rangle+Z_{i}\Omega_{i\alpha}) \delta_{i j}
\delta_{\alpha \beta}\\ \nonumber + t^{\alpha \beta}_{i
j}(1-\delta_{ij})] |i\alpha\rangle \langle j\beta| \left[
\begin{array}{cc}
1 & 0\\
0 & 1
\end{array}\right]
%\end{eqnarray}
%\end{equation}
\\
%\begin{eqnarray}
%\begin{equation}
\nonumber H_{exch}=\sum_{i,\alpha}(-{1\over2}
J_{i\alpha}\mu_{i_\alpha}) |i\alpha\rangle \langle i\alpha|\left[
\begin{array}{cc}
\cos \theta_i & e^{-i\phi_i}\sin \theta_i\\
e^{i\phi_i}\sin \theta_i & - \cos \theta_i
\end{array}\right]
%\end{equation}
\end{eqnarray}

$H_{band}$ contains both the non-diagonal matrix elements (hopping
integrals, $t^{\alpha \beta}_{i j}$, between orbitals $\alpha$ and
$\beta$ of different sites $i$ and $j$, which are assumed to be
spin independent) and the spin-independent part of the diagonal
matrix elements ($\epsilon ^{0}_{i\alpha } +U_{i\alpha,j\beta}
\langle \hat{n}_{jb}\rangle+Z_{i}\Omega_{i\alpha}$), being the sum
of the local level $\epsilon ^{0}_{i\alpha }$, the electrostatic
level shift $U_{i\alpha,j\beta}\langle \hat{n}_{jb}\rangle$
accounting for the charge variations parameterized by the Coulomb
integral $U_{i\alpha,j\beta}$ and the crystal field potential
$Z_i\Omega_{i\alpha}$, where $Z_i$ is the local atomic
coordination of site $i$. $H_{exch}$ describes the magnetic part,
through the exchange parameter $J_{i\alpha}$ multiplied by the
magnitude of the local magnetic moment, $\mu_{i\alpha}$, whose
direction is given by the angles $(\theta_i,\phi_i)$ in the
spin-rotation matrix. The spin-orbit contribution is not
considered here. The Hamiltonian has been parametrized to DFT
TB-LMTO calculations of thin Mn films supported on ideal Fe(001)
\cite{physletA} in order to take into account the effects of
surface, interface and bulk.The transferability of the parametrization has
been checked in systems of 6 and 7 Mn ML on Fe(001), comparing the
results with those obtained with the TB-LMTO method
\cite{physletA}, finding similar values for the energy differences
and magnetic moments. 

\begin{figure*}
%\begin{center}
\includegraphics*[width=157mm]{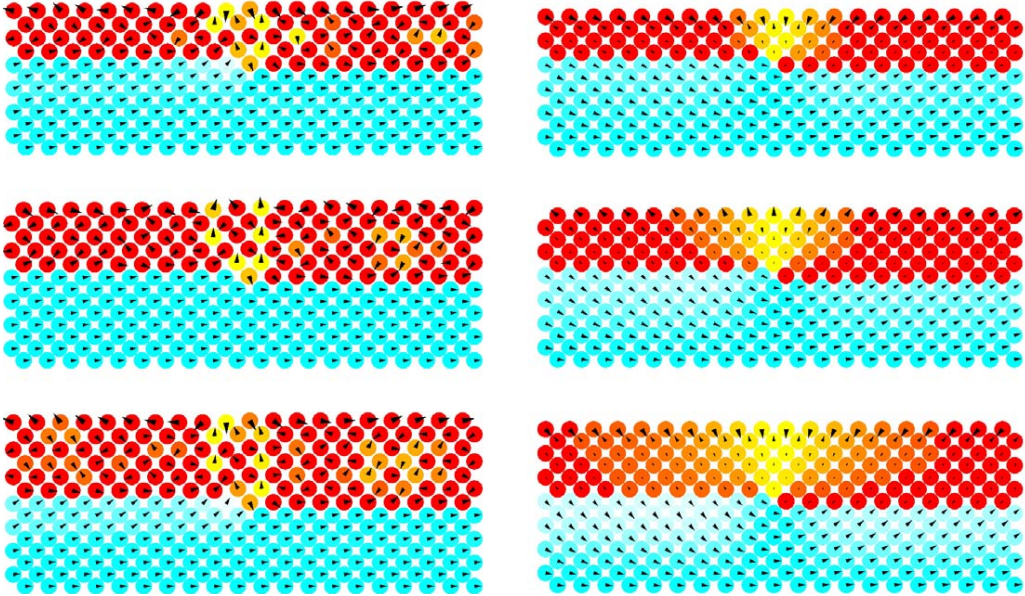}
\caption{(color online) Result of the calculation performed for
4-5, 5-6 and 6-7 monolayers of Mn (left panel) and Cr (right
panel) supported on an Fe(001) substrate with a monoatomic step
(only a portion of the calculated systems is represented). The
arrows indicates the orientation of the individual magnetic moment
and its size is proportional to absolute value of the local
magnetic moment (average values of the magnetic moments far from
the step: $\mu^{bulk}_{Fe}=2.32 \mu_{B}$, $\mu^{surf}_{Mn}=3.84
\mu_{B}$, $\mu^{surf}_{Cr}=3.07 \mu_{B}$, $\mu^{inter}_{Mn}=2.85
\mu_{B}$, $\mu^{inter}_{Cr}=1.13 \mu_{B}$). The color of the
circles is given by the absolute value of the cosine of the angle
of the magnetic moment. For Mn and Cr red means a LAF arrangement
and yellow indicates the magnetic moment at 90$^o$. For the
Fe(001) substrate the blue color indicates a ferromagnetic
arrangement and light blue a deviation from that.}
\label{figure4}
%\end{center}
\end{figure*}

We have simulated the systems represented in Fig. \ref{figure1}.
Despite the fact of preserving the symmetry in the axis parallel
to the step, that can be justified by the length of the observed
steps in the system, we have about 40000 atoms and up to 600
inequivalent sites to describe the semi-infinite system. We have
considered coverages of 4-5, 5-6 and 6-7 Mn ML. In all cases, a
small step of about 0.02 nm at the surface is present (as
experimentally observed). In Fig. \ref{figure4} (left panel) we
illustrate the noncollinear magnetic moment distributions obtained
in the calculations (only a portion of the semi-infinite system is
shown). To understand the origin of the noncollinear magnetic
arrangement, we have also performed TB collinear restricted
calculations (not shown), obtaining different magnetic
arrangements, the least energetic one displaying the LAF order at
both sides far from the step, while just over the step, due to the
impossibility to reach in the whole system the LAF order together
with the parallel coupling at the interface, magnetically
frustrated atoms were present. Previous studies in Cr/Fe
interfaces with steps \cite{nocolcr} have shown that this type of
frustrations in the collinear framework lead the system to
drastically reduce the magnetic moments in the frustrated region.
However, this moment reduction in the case of Mn is much less
noticeable, which is consistent with the fact that the magnetic
couplings between Mn layers are weaker and less defined than in Cr
for which the LAF solution is the only one present \cite{nocolcr}.
The noncollinear magnetic solutions shown in Fig. \ref{figure4}
are energetically more stable than the collinear ones in all cases
and reproduce the experimentally observed magnetic contrast at the
surface between both sides of the step. This is due to the
propagation of the magnetic frustration originated at the
interface step. Far from the step, the system tends to preserve
the magnetic couplings found in the ideal Mn films on Fe(001)
\cite{physletA}, that is the LAF with parallel Fe-Mn coupling at
the interface and antiparallel coupling between Mn surface and
subsurface layers. Moreover, the tendency of Mn to couple both
parallel and antiparallel in the central layers of the ideal Mn
films on Fe(001) is also reflected in the noncollinear
arrangements shown in Fig. \ref{figure4}

\begin{figure*}
%\begin{center}
\includegraphics*[angle=270.0, width=157mm]{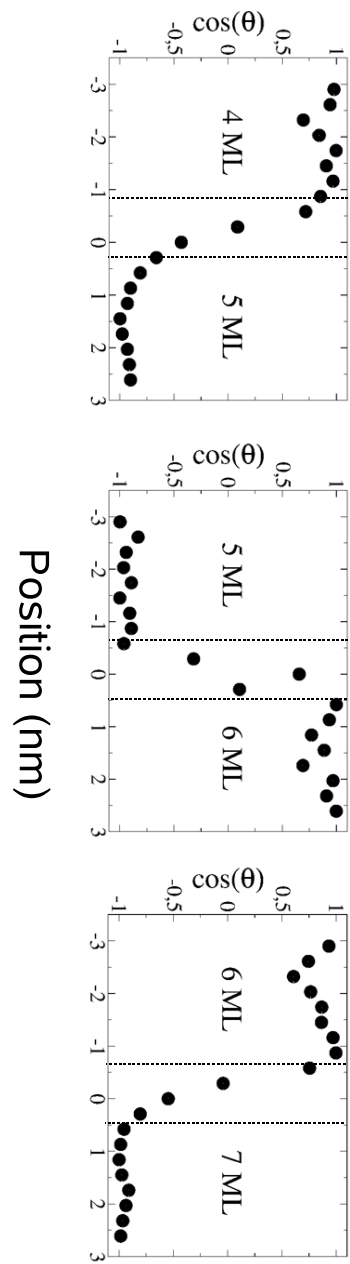}
\caption{Cosine of the angle of the magnetic surface
moments with respect to the bulk obtained in the theoretical
calculations shown in Fig. \ref{figure4} for 4-5, 5-6 and 6-7 ML. Vertical
dashed lines specify the size of the domain wall in each case.}
\label{figure5}
%\end{center}
\end{figure*}

\begin{figure*}
%\begin{center}
\includegraphics*[angle=270.0, width=80mm]{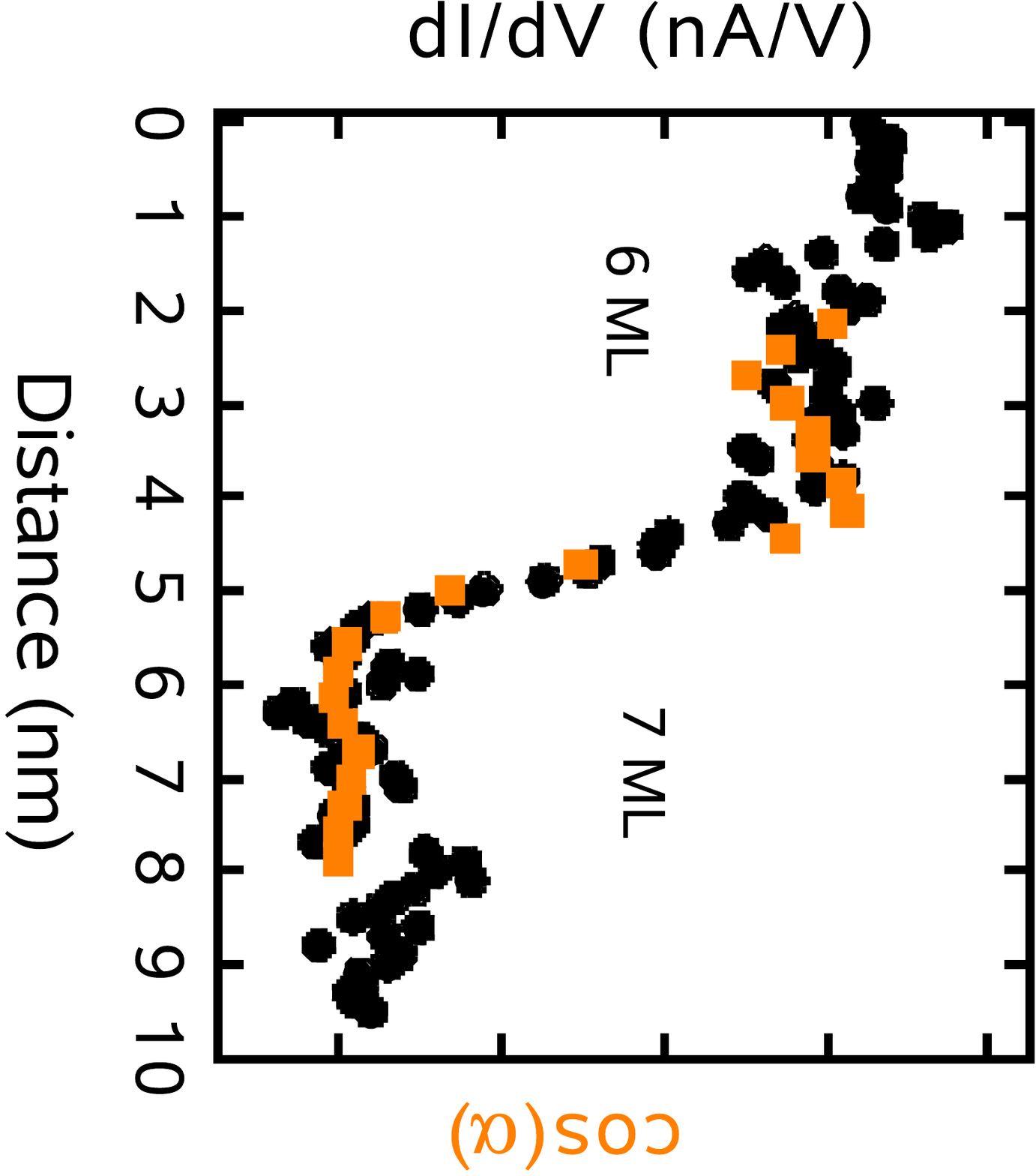}
\caption{(color online) (black dots) Average experimental
profile across the hidden step measured inside the white box shown
in Fig. \ref{figure1}(b). (orange squares) Cosine of the angle of the magnetic surface
moments with respect to the bulk obtained in the theoretical
calculation shown in Fig. \ref{figure4} for 6-7 ML.}
\label{figure6}
%\end{center}
\end{figure*}

Let us now come to the analysis of the domain-wall evolution as a
function of the Mn coverage, in particular the width of the
domain-wall. Experimentally, we have found a domain-wall width of
about 2-4 lattice parameters independently of the Mn coverage. Our
theoretical results (Fig. \ref{figure4}) are also consistent with
the experimental finding, as it can be deduced from Figs.
\ref{figure5} and \ref{figure6} where the cosine of the angle of the local
surface magnetic moment with respect to the bulk are plotted
through the step (red squares). The central Mn layers decouple the
surface from the interface in a large
extent, and make these systems to behave in a
radical different way from what would be expected if one assumes
for Mn the magnetic coupling of a typical antiferromagnet like Cr
and the associated magnetic frustrations when interfaced with
stepped Fe. In fact, we have already indicated before that Mn and
Cr have different magnetic behavior when deposited on an
atomically flat Fe(001) surface. The magnetic decoupling of the
surface layers from the deeper layers coupled to the substrate has
been also proposed recently by Hafner and Spi\v{s}\'ak
\cite{hafner}. This provides further support to our results. 
For coverages larger enough it is expected that Mn will undergo
structural transition \cite{passamani} with no
magnetic domains at the surface. Our
results contrast, however, with those obtained by Schlickum {\it et al}.
\cite{schlickum}. These authors found an increase of the
domain-wall width as increasing Mn coverage. Growth conditions are
very important for the final structure. In these experiments we
grew the Mn films about ten times faster than Schlickum {\it et al}.
\cite{schlickum}. The faster deposition rate for the 
same sample temperature may lead
to a smaller level of intermixing between Mn and 
Fe and therefore to a different
magnetic structure. Furthermore Schlickum {\it et al}. \cite{schlickum}
explained their results using a simple Heisenberg model for
localized magnetism \cite{schlickum} assuming that Mn behaves like
Cr and considering the same exchange parameter for Mn-Fe and
Mn-Mn.

In order to further illustrate the different magnetic behavior of Mn and
Cr, we have performed the same type of calculations in similar
systems, but with Cr instead of Mn deposited on the stepped Fe
substrate and the corresponding interatomic distances (Fig. \ref{figure4},
right panel). The difference between the noncollinear arrangements
in Mn/Fe and Cr/Fe is clear. The strong antiferromagnetic
character of Cr leads to strong magnetic frustrations that are
propagated towards the surface occupying a region in which the
best compromise to the antiparallel Cr-Cr and Cr-Fe couplings is
achieved. As a result, the domain-wall width increases with
increasing Cr coverage and the frustration is partially released
in the Fe(001) substrate.

\section{Conclusions}
In conclusion, we have investigated the magnetic structure on thin
Mn films grown on a Fe(001) surface with a monoatomic step. In our
SP-STM images we found a change in the magnetic contrast when
crossing one of those steps due to the change of the Mn thickness.
We have found that the width of the domain wall around the substrate steps does not depend on
the thickness, at least for coverages up to 7 Mn overlayers, and it is about 2
lattice parameters. This is due to the weakly defined magnetic coupling at the
central Mn layers that decouple the surface from the interface in a large extent. We
compare our findings for the Mn films with the behavior of Cr
films. In contrast to Mn, ideal Cr films have strong antiferromagnetic couplings and
larger range order than in Mn, up to the point that only one magnetic arrangement in the system
has been found in the collinear framework, corresponding to a LAF configuration. This strong
coupling produces a high amount of frustration on steps and
as a consequence in the case of Cr the domain wall width increases with the coverage.

Financial support by the Ministerio de Ciencia y Tecnolog\'{\i}a
and Ministerio de Educaci\'on y Ciencia through project numbers
FIS2004-01206 and MAT2005-03415, Junta de Castilla y Le\'on (VA068A06) and INTAS (03-51-4778) is gratefully acknowledge. HvK
thanks Instituto Universitario "Nicol\'as Cabrera" for a visitors
grant.

\end {document}